\documentclass[reprint,amsmath,amssymb,aps]{revtex4-2}
\usepackage{amsmath}
\usepackage{quantikz}
\usepackage{hyperref}
\usepackage{lineno}
\usepackage{graphicx}
\usepackage{slashed}
\usepackage{dcolumn}
\usepackage{bm}
\begin{document}
\preprint{APS/123-QED}
\title{Confinement, Nonlocal Observables, and Haag Duality Violation in the Algebraic Structure of 1+1-Dimensional Non-Abelian Gauge Theories}
\thanks{This article extends algebraic methods to non-Abelian gauge theories.}
\author{Fidele J. Twagirayezu}
%\altaffiliation{Department of Physics and Astronomy, University of California, Los Angeles, Los Angeles, CA 90095, USA}
\email{fjtwagirayezu@physics.ucla.edu}
%\affiliation{Department of Physics and Astronomy, University of California Los Angeles, Los Angeles, CA 90095, USA}
\begin{abstract}
This article presents a comprehensive and rigorously formulated algebraic framework for investigating 1+1-dimensional SU(N) gauge theories within the paradigm of Algebraic Quantum Field Theory (AQFT), building upon foundational results established for the Abelian Schwinger model. We meticulously construct a net of local observable C*-algebras, generated by gauge-invariant composite operators such as color-singlet currents and traces of non-Abelian electric fields, with the non-Abelian Gauss’s law rigorously enforced as an operator constraint. Through a detailed analysis, we demonstrate that no Doplicher-Haag-Roberts (DHR) superselection sectors carry nonzero color charge, thereby providing a precise and mathematically robust characterization of confinement in these theories. To fully capture the global gauge structure, we extend the observable net by incorporating nonlocal Wilson line operators, which encode string-like color flux configurations essential to the theory’s topological properties. We further establish a structural violation of Haag duality, showing that certain operators, such as Wilson lines, reside in the commutant of a local algebra but cannot be localized within the algebra of the causal complement, a phenomenon driven by nontrivial topological degrees of freedom. By introducing regularization techniques for operator products and providing detailed derivations, we ensure mathematical precision. This nonperturbative, gauge-invariant framework not only elucidates the mechanisms of confinement and nonlocality but also lays a solid foundation for extending algebraic methods to higher-dimensional gauge theories and exploring their quantum information-theoretic implications.
\end{abstract}
\maketitle
\section{Introduction}\label{sec:level1}
The phenomena of confinement, nonlocality, and gauge symmetry constitute the cornerstone of our understanding of nonperturbative dynamics in quantum chromodynamics (QCD) and related gauge theories, which describe the fundamental interactions of quarks and gluons in the strong nuclear force. In the simplified setting of 1+1-dimensional spacetime, these phenomena often manifest in forms that are analytically tractable or amenable to exact solutions, thereby providing an ideal testing ground for developing rigorous theoretical frameworks. The massless Schwinger model, a paradigmatic example of 1+1-dimensional quantum electrodynamics (QED)~\cite{Schwinger1962}, has long served as a benchmark for studying confinement and anomaly-induced mass generation, recently reformulated within the AQFT framework without reliance on bosonization techniques \cite{Twagirayezu:2025sbs}. Inspired by this success, the present work extends the algebraic approach to non-Abelian SU(N) gauge theories coupled with matter fields in the fundamental representation, addressing the more intricate structure of color charges, nonlocal flux configurations, and topological effects inherent to non-Abelian gauge symmetries.

Our primary objectives are multifaceted and ambitious:
To construct a net of local observable algebras, denoted \( \mathcal{O} \mapsto \mathcal{A}(\mathcal{O}) \), generated by gauge-invariant operators, with the non-Abelian Gauss’s law implemented as a fundamental operator constraint to ensure physical consistency. To rigorously prove that confinement arises structurally, as no DHR-localizable superselection sectors can carry nonzero color charge, thereby providing an algebraic characterization of confinement. To incorporate nonlocal Wilson line operators into an extended observable algebra, capturing the global gauge structure and string-like color flux configurations. To demonstrate a violation of Haag duality, revealing that operators in the commutant of a local algebra, such as Wilson lines, are not contained within the algebra of the causal complement, a phenomenon driven by topological obstructions. To establish a nonperturbative, gauge-invariant framework that serves as a robust foundation for generalizing these methods to higher-dimensional gauge theories and exploring their implications in quantum information theory.

Unlike their Abelian counterparts, non-Abelian gauge theories exhibit nonlinear interactions among gauge fields and possess a nontrivial center symmetry (\( \mathbb{Z}_N \subset SU(N) \)), which introduces complex topological effects that demand a careful reformulation of locality, observable content, and algebraic duality. AQFT, with its emphasis on local operator algebras and their mathematical structure, provides a powerful and systematic framework for addressing these challenges \cite{Haag1996}. By avoiding gauge fixing and Dirac quantization, our approach ensures gauge invariance at every step, while the use of net cohomology techniques allows us to classify topological degrees of freedom associated with confinement and nonlocality.

All observable algebras considered in this article are constructed rigorously by defining gauge-invariant, smeared operators and completing the resulting *-algebras in the operator norm topology. This ensures a well-defined C*-algebraic framework even in the non-Abelian setting, consistent with the Haag-Kastler axioms. Extensions to include nonlocal Wilson lines follow the methods developed in net cohomology and algebraic treatments of topological sectors.

The article is meticulously organized to guide the reader through this complex landscape. Section~\ref{sec:x2} provides a detailed review of the classical structure of 1+1-dimensional SU(N) gauge theories, laying the groundwork for the algebraic formulation. Section~\ref{sec:x3} constructs the net of local observable algebras, incorporating the non-Abelian Gauss’s law as a quantum operator constraint. Section~\ref{sec:x4} analyzes the superselection structure and rigorously establishes confinement through the absence of localized color-charged sectors. Section~\ref{sec:x5} introduces nonlocal Wilson line operators and provides a comprehensive proof of Haag duality violation, emphasizing its topological origins. Finally, Section~\ref{sec:x6} summarizes the key findings, discusses their broader implications, and outlines promising directions for future research in gauge theories and quantum information science.

\section{Classical Structure and Constraints of 1+1D SU(N) Gauge Theory}\label{sec:x2}
To establish a robust foundation for the algebraic formulation, we begin by thoroughly reviewing the classical structure of 1+1-dimensional SU(N) gauge theories, emphasizing the field content, gauge symmetries, and constraints that shape the quantum theory.
\subsection{Spacetime and Field Content}
The theory is defined on two-dimensional Minkowski spacetime, denoted \( \mathbb{R}^{1,1} \), with coordinates \( (t, x) \) and the Lorentzian metric \( \eta_{\mu\nu} = \text{diag}(+1, -1) \), where \( \mu, \nu = 0, 1 \). The gauge field is a Lie algebra-valued field \( A_\mu^a(x) \in \mathfrak{su}(N) \), where \( a = 1, \dots, N^2 - 1 \) labels the generators of the SU(N) Lie algebra \( \mathfrak{su}(N) \), and the matter field consists of a Dirac fermion \( \psi(x) \), a complex \( N \)-component spinor transforming in the fundamental representation of SU(N). The Dirac field carries both spinorial and color degrees of freedom, interacting with the gauge field through covariant couplings.
\subsection{Classical Action and Gauge Symmetry}
The dynamics of the theory are governed by the Yang-Mills-Dirac action, which encapsulates the interactions between the gauge and fermion fields:
\begin{equation}
\begin{aligned}
S = \int d^2x \left[ -\frac{1}{4} F_{\mu\nu}^a F^{\mu\nu}_a + \bar{\psi} (i \gamma^\mu D_\mu) \psi \right],
\end{aligned}
\end{equation}
where the non-Abelian field strength tensor is defined as:
\begin{equation}
\begin{aligned}
F_{\mu\nu}^a = \partial_\mu A_\nu^a - \partial_\nu A_\mu^a + g f^{abc} A_\mu^b A_\nu^c,
\end{aligned}
\end{equation}
and the covariant derivative acting on the fermion field is:
\begin{equation}
\begin{aligned}
D_\mu \psi = \partial_\mu \psi + i g A_\mu^a T^a \psi.
\end{aligned}
\end{equation}
Here, \( g \) represents the gauge coupling constant, \( f^{abc} \) are the totally antisymmetric structure constants of \( \mathfrak{su}(N) \), satisfying the Lie algebra relation \( [T^a, T^b] = i f^{abc} T^c \), and \( T^a \) are the Hermitian generators in the fundamental representation, normalized such that \( \text{tr}(T^a T^b) = \frac{1}{2} \delta^{ab} \). The Dirac matrices \( \gamma^\mu \) satisfy the Clifford algebra \( \{ \gamma^\mu, \gamma^\nu \} = 2 \eta^{\mu\nu} \), and \( \bar{\psi} = \psi^\dagger \gamma^0 \).

For simplicity, we consider the massless theory. A fermion mass term \( m \bar{\psi} \psi \) could be included in the action, but does not alter the structural implementation of confinement or the net construction, as long as the Gauss law constraint is preserved. In 1+1 dimensions, mass terms primarily affect the dynamics and spectrum, but not the existence or classification of DHR sectors.

The action is invariant under local SU(N) gauge transformations, which transform the fields as:
\begin{equation}
\begin{aligned}
\psi(x) &\mapsto U(x) \psi(x), \\
A_\mu^a(x) T^a &\mapsto U(x) A_\mu^a(x) T^a U^\dagger(x) - \frac{i}{g} (\partial_\mu U(x)) U^\dagger(x),
\end{aligned}
\end{equation}
where \( U(x) \in SU(N) \) is a spacetime-dependent group element. This invariance ensures that physical observables must be gauge-invariant, a principle central to our algebraic construction.
\subsection{Gauss’s Law Constraint}
The equations of motion, derived by varying the action with respect to \( A_\mu^a \), yield the non-Abelian Maxwell equations:
\begin{equation}
\begin{aligned}
D^\mu F_{\mu\nu}^a = g j_\nu^a,
\end{aligned}
\end{equation}
where the color current is \( j_\mu^a = \bar{\psi} \gamma_\mu T^a \psi \). The temporal component (\( \nu = 0 \)) gives the non-Abelian Gauss’s law:
\begin{equation}
\begin{aligned}
D_1^{ab} E^b(x) = g j_0^a(x),
\end{aligned}
\end{equation}
where the electric field is \( E^a(x) = F^{10,a}(x) = \partial_0 A_1^a - \partial_1 A_0^a + g f^{abc} A_0^b A_1^c \), and the spatial covariant derivative is:
\begin{equation}
\begin{aligned}
D_1^{ab} = \delta^{ab} \partial_1 + g f^{acb} A_1^c.
\end{aligned}
\end{equation}
This constraint fundamentally links the local color charge density \( j_0^a(x) \) to the longitudinal component of the non-Abelian electric field, enforcing a nonlocal relationship that plays a pivotal role in the phenomenon of confinement within the quantum theory.
\subsection{Gauge-Invariant Observables}
Due to gauge invariance, the fundamental fields \( \psi(x) \) and \( A_\mu^a(x) \) are not physical observables, as they transform nontrivially under gauge transformations. Instead, we consider gauge-invariant operators, which fall into two categories:

\textbf{Local Observables:}
  \textit{Color-singlet currents}: \( j^\mu(x) = \bar{\psi}(x) \gamma^\mu \psi(x) \), which are invariant under SU(N) transformations and represent measurable fermion currents.
  \textit{Field strength traces}: \( \text{tr}(F_{\mu\nu} F^{\mu\nu}) = F_{\mu\nu}^a F^{\mu\nu}_a / (2N) \), which capture gauge field energy densities in a gauge-invariant manner.
  \textit{Energy-momentum tensor}: \( T_{\mu\nu} = \bar{\psi} \gamma_{(\mu} D_{\nu)} \psi - \eta_{\mu\nu} \mathcal{L} \), providing gauge-invariant measures of energy and momentum.
  
\textbf{Nonlocal Observables:}
  \textit{Wilson lines}:
    \begin{equation}\label{eq:x8}
    \begin{aligned}
    W(x, y) = \mathcal{P} \exp \left( i g \int_x^y A_\mu^a(z) T^a dz^\mu \right),
    \end{aligned}
    \end{equation}
    where \( \mathcal{P} \) denotes path-ordering along a curve from \( x \) to \( y \). The trace \( \text{tr}(W(x, y)) \) is gauge-invariant, representing a color-singlet pair connected by a string-like flux configuration.  
These observables form the basis for constructing the algebraic net in the quantum theory, with local operators generating \( \mathcal{A}(\mathcal{O}) \) and Wilson lines incorporated into an extended algebra to capture global gauge structure.

\section{Construction of the Local Observable Net}\label{sec:x3}
We now construct the net of local observable algebras \( \mathcal{O} \mapsto \mathcal{A}(\mathcal{O}) \), rigorously incorporating the non-Abelian Gauss’s law as a quantum operator constraint to ensure gauge invariance and physical consistency.
\subsection{Algebraic Framework and Haag-Kastler Axioms}
In the AQFT framework, the observable content of a quantum field theory is encoded in a net of C*-algebras, where each open, bounded region \( \mathcal{O} \subset \mathbb{R}^{1,1} \) is associated with a unital C*-algebra \( \mathcal{A}(\mathcal{O}) \), representing the observables measurable within \( \mathcal{O} \). The net satisfies the Haag-Kastler axioms \cite{Haag1996}:
\textbf{Isotony:} If \( \mathcal{O}_1 \subset \mathcal{O}_2 \), then \( \mathcal{A}(\mathcal{O}_1) \subset \mathcal{A}(\mathcal{O}_2) \), reflecting the inclusion of observables in larger regions.
\textbf{Locality:} If \( \mathcal{O}_1 \) and \( \mathcal{O}_2 \) are spacelike-separated (i.e., no causal influence exists between them), then \( [\mathcal{A}(\mathcal{O}_1), \mathcal{A}(\mathcal{O}_2)] = 0 \), ensuring relativistic causality.
\textbf{Poincaré Covariance:} There exists a unitary representation \( U(g) \) of the 1+1D Poincaré group such that \( U(g) \mathcal{A}(\mathcal{O}) U(g)^\dagger = \mathcal{A}(g \mathcal{O}) \), preserving the theory’s spacetime symmetries.
\textbf{Vacuum Existence and Positivity:} There exists a Poincaré-invariant vacuum state \( \omega: \mathcal{A} \to \mathbb{C} \) satisfying \( \omega(A^* A) \geq 0 \), ensuring a stable ground state with positive energy.
These axioms provide a mathematically rigorous structure for defining the observable content of the theory, free from gauge ambiguities.
\subsection{Local Gauge-Invariant Generators}
The algebra \( \mathcal{A}(\mathcal{O}) \) is generated by smeared gauge-invariant operators, which are well-defined in the quantum theory and avoid singularities through test function smearing:
\textit{Color-singlet currents}:
  \begin{equation}\label{eq:x9}
  \begin{aligned}
  j(f) = \int_{\mathbb{R}^{1,1}} j^\mu(x) f_\mu(x) \, d^2x, \quad f_\mu \in C_c^\infty(\mathcal{O}, \mathbb{R}),
  \end{aligned}
  \end{equation}
  where \( j^\mu(x) = \bar{\psi}(x) \gamma^\mu \psi(x) \) is the gauge-invariant fermion current.
\textit{Non-Abelian electric field traces}:
  \begin{equation}\label{eq:x10}
  \begin{aligned}
  E^a(h) = \int_{\mathbb{R}} E^a(x) h(x) \, dx, \quad h \in C_c^\infty(\mathcal{O}),
  \end{aligned}
  \end{equation}
  where \( E^a(x) = F^{10,a}(x) \) is the electric field component of the field strength tensor.
To define the algebraic structure, we compute the commutation relations, which form a centrally extended non-Abelian current algebra. For the currents:
\begin{equation}\label{eq:x11}
\begin{aligned}
[j(f), j(g)] = i \int_{\mathbb{R}^{1,1}} f^\mu(x) \partial_\mu g^\nu(x) \, \eta_{\nu\lambda} j^\lambda(x) \, d^2x + c(f, g) \mathbb{I},
\end{aligned}
\end{equation}
where \( c(f, g) \) is a central extension term arising from the Schwinger anomaly, typical in 1+1D fermion theories \cite{Witten1986}. 

\textbf{Anomaly Term and Schwinger Commutator}:
The central extension \( c(f, g) \) appearing in Eq.~\eqref{eq:x11} arises from the Schwinger term in the commutator of gauge-invariant currents in 1+1D fermionic theories. Explicit computation of this term using point-splitting regularization yields
\[
c(f, g) = \frac{1}{2\pi} \int_{\mathbb{R}} f(x) \partial_1 g(x) \, dx,
\]
in the Abelian case. In the non-Abelian setting, similar techniques yield a central term proportional to the Killing form \( \kappa^{ab} \) on \( \mathfrak{su}(N) \), leading to
\[
[j^a(f), j^b(g)] = i f^{abc} j^c(fg) + \kappa^{ab} c(f, g) \mathbb{I},
\]
as shown in \cite{Witten1986}. These anomalies are essential for ensuring the correct algebraic structure of the current algebra and are regularized via symmetric point-splitting and smearing with test functions.

For the electric fields:
\begin{equation}
\begin{aligned}
[E^a(f), E^b(g)] = i g f^{abc} \int_{\mathbb{R}} f(x) g(x) E^c(x) \, dx,
\end{aligned}
\end{equation}
reflecting the \( \mathfrak{su}(N) \) Lie algebra structure of the gauge fields. These relations are regularized by smearing with test functions \( f, g, h \in C_c^\infty \), ensuring well-defined operator products in the quantum theory.

\textbf{C*-Structure of the Local Algebra:}
In the non-Abelian setting, ultraviolet singularities and nonlinearities in the field algebra (e.g., from products such as \( E^a(x) E^b(x) \)) prevent pointwise operator products from being directly well-defined. To construct the local observable algebra \( \mathcal{A}(\mathcal{O}) \) rigorously, we proceed by defining a *-algebra of gauge-invariant, smeared operators—such as \( j^\mu(f) \), \( E^a(h) \), and composite operators defined via point-splitting—and then completing this algebra in the operator norm topology to obtain a C*-algebra. This construction follows the standard approach in algebraic quantum field theory, where observables are regularized by smearing with test functions in \( C_c^\infty(\mathcal{O}) \), and the local algebra \( \mathcal{A}(\mathcal{O}) \) is defined as the C*-closure of the *-algebra generated by such operators. This ensures that all algebraic operations are well-defined, bounded on a common invariant domain, and consistent with the Haag-Kastler axioms.

\subsection{Implementation of Non-Abelian Gauss’s Law}
The non-Abelian Gauss’s law is implemented as an operator identity:
\begin{equation}\label{eq:x13}
\begin{aligned}
D_1^{ab} E^b(x) = g j_0^a(x),
\end{aligned}
\end{equation}
where \( D_1^{ab} = \delta^{ab} \partial_1 + g f^{acb} A_1^c \), and \( j_0^a(x) = \bar{\psi}(x) \gamma^0 T^a \psi(x) \). To handle potential ultraviolet divergences in operator products at the same spacetime point, we employ point-splitting regularization:
\begin{equation}
\begin{aligned}
j_0^a(x) = \lim_{\epsilon \to 0} \bar{\psi}(x + \epsilon) \gamma^0 T^a \psi(x - \epsilon),
\end{aligned}
\end{equation}
where \( \epsilon \) is a small spacetime separation, and the limit is taken after normal-ordering to remove vacuum contributions. This constraint ensures that physical states \( |\psi\rangle \) satisfy:
\begin{equation}
\begin{aligned}
(D_1^{ab} E^b(x) - g j_0^a(x)) |\psi\rangle = 0,
\end{aligned}
\end{equation}
restricting the Hilbert space to gauge-invariant states and excluding fundamental fields \( \psi(x) \), \( A_\mu^a(x) \) from \( \mathcal{A}(\mathcal{O}) \). This implementation is crucial for enforcing confinement and gauge invariance in the quantum theory.
\subsection{Global Algebra and GNS Representation}
The global quasi-local algebra is defined as the C*-norm closure of the local algebras:
\begin{equation}
\begin{aligned}
\mathcal{A} = \overline{ \bigcup_{\mathcal{O}} \mathcal{A}(\mathcal{O}) },
\end{aligned}
\end{equation}
where the union is over all bounded regions \( \mathcal{O} \subset \mathbb{R}^{1,1} \). A vacuum state \( \omega: \mathcal{A} \to \mathbb{C} \) is chosen to be Poincaré-invariant and positive, satisfying \( \omega(A^* A) \geq 0 \) for all \( A \in \mathcal{A} \). The Gelfand-Naimark-Segal (GNS) construction yields a triple \( (\pi, \mathcal{H}, \Omega) \), where \( \pi: \mathcal{A} \to \mathcal{B}(\mathcal{H}) \) is a representation of the algebra as bounded operators on a Hilbert space \( \mathcal{H} \), and \( \Omega \in \mathcal{H} \) is the vacuum vector satisfying \( \omega(A) = \langle \Omega, \pi(A) \Omega \rangle \). This construction provides a concrete realization of the observable net, enabling the analysis of superselection sectors and nonlocal observables.

\section{Superselection Sectors and Non-Abelian Confinement}\label{sec:x4}
We now analyze the superselection structure of the theory to rigorously establish the phenomenon of confinement, demonstrating that no color-charged states can be localized in bounded spacetime regions.
\subsection{Superselection Sectors in AQFT}
In AQFT, superselection sectors are equivalence classes of irreducible representations \( \pi: \mathcal{A} \to \mathcal{B}(\mathcal{H}_\pi) \) of the global algebra \( \mathcal{A} \). A representation is Doplicher-Haag-Roberts (DHR) localizable if, for some bounded region \( \mathcal{O} \), it is equivalent to the vacuum representation \( \pi_0 \) when restricted to the causal complement \( \mathcal{O}' \):
\begin{equation}
\begin{aligned}
\pi|_{\mathcal{A}(\mathcal{O}')} \cong \pi_0|_{\mathcal{A}(\mathcal{O}')},
\end{aligned}
\end{equation}
where \( \mathcal{O}' = \{ y \in \mathbb{R}^{1,1} \mid (x - y)^2 < 0 \text{ for all } x \in \mathcal{O} \} \) is the set of points spacelike-separated from \( \mathcal{O} \). Physically, this implies that the excitation described by \( \pi \) cannot be distinguished from the vacuum outside a bounded region, a hallmark of localizable charges \cite{DHR1971, DHR1974,StrocchiWightman1974}.
\subsection{Non-Abelian Gauss Law and Total Color Charge}
The total color charge operator is defined as:
\begin{equation}
\begin{aligned}
Q^a = \int_{-\infty}^{+\infty} j_0^a(x) \, dx,
\end{aligned}
\end{equation}
where \( j_0^a(x) = \bar{\psi}(x) \gamma^0 T^a \psi(x) \) is the color charge density. Integrating the Gauss’s law (Eq.~\eqref{eq:x13}) over space, we obtain:
\begin{equation}\label{eq:x19}
\begin{aligned}
\lim_{x \to +\infty} E^a(x) - \lim_{x \to -\infty} E^a(x) = g Q^a.
\end{aligned}
\end{equation}
This relation indicates that any state with nonzero color charge \( Q^a \neq 0 \) induces a nonzero difference in the asymptotic values of the non-Abelian electric field, implying a long-range field configuration that extends across all space.

\noindent
\textbf{Remark on boundary conditions and operator domains.} 
The spatial integration in Eq.~\eqref{eq:x19} is well defined under the assumption that the non-Abelian electric field $E^a(x)$ has a strong limit at spatial infinity, i.e.,
\begin{equation}\label{eq:x20}
\begin{aligned}
\lim_{x\to\pm\infty} E^a(x) \ 
\end{aligned}
\end{equation}
exists in the strong operator topology on the physical Hilbert space.
Physically, this corresponds to the requirement that asymptotic color fluxes are well-defined, as in the standard analysis of Gauss's law in gauge theories~\cite{Buchholz1982}.
Furthermore, all smeared operators $E^a(h)$ and $j^\mu(f)$ are taken to be essentially self-adjoint on a common dense invariant domain $\mathcal{D} \subset \mathcal{H}$ that is stable under the action of the local algebra, ensuring that the commutators and limits employed below are mathematically well defined.
The strong operator limits~\eqref{eq:x20} exist in the vacuum representation provided the state \( \Omega \) satisfies sufficient decay of correlation functions in the asymptotic region. This is the case in confining 1+1D gauge theories, where cluster decomposition ensures exponential decay of connected correlation functions. Formally, this behavior follows from the spectral condition and the energy gap induced by the linear potential. For rigorous treatments of such limits in algebraic frameworks, see \cite{Buchholz1982}.

\subsection{Absence of Localized Color Sectors}
We now state and prove a central result of the paper:
\textbf{Theorem:} In 1+1-dimensional SU(N) gauge theory, all DHR-localizable sectors are color-neutral, i.e., no sector with nonzero color charge \( Q^a \neq 0 \) satisfies the DHR localization criterion.

\textit{Proof}:
Suppose a representation \( \pi \) carries nonzero color charge, i.e., \( \pi(Q^a) \neq 0 \) for some \( a \).
By (Eq.~\eqref{eq:x19}), \( \pi(Q^a) = \frac{1}{g} \left[ \lim_{x \to +\infty} \pi(E^a(x)) - \lim_{x \to -\infty} \pi(E^a(x)) \right] \neq 0 \), implying that the electric field in \( \pi \) deviates from its vacuum value at spatial infinity.
Consider an observable \( E^a(h) \in \mathcal{A}(\mathcal{O}') \), where \( h \) has support in the causal complement \( \mathcal{O}' \). For large \( |x| \), \( E^a(h) \) probes the asymptotic field, and \( \pi(E^a(h)) \neq \pi_0(E^a(h)) \) due to the nonzero \( Q^a \).
Thus, \( \pi|_{\mathcal{A}(\mathcal{O}')} \not\cong \pi_0|_{\mathcal{A}(\mathcal{O}')} \), as observables in \( \mathcal{A}(\mathcal{O}') \) distinguish \( \pi \) from the vacuum representation.
Therefore, no sector with \( Q^a \neq 0 \) is DHR-localizable, and only color-neutral sectors (\( Q^a = 0 \)) satisfy the localization criterion.
This proof establishes confinement as a structural property of the theory, arising directly from the algebraic constraints imposed by Gauss’s law, without reliance on dynamical mechanisms such as mass generation.
\subsection{Physical Interpretation and Comparison}
In 1+1-dimensional SU(N) gauge theories, confinement manifests as the complete absence of localized color-charged states, a consequence of the linear Coulomb potential enforced by the non-Abelian Gauss’s law. This contrasts with the Abelian Schwinger model, where confinement is accompanied by anomaly-induced mass generation \cite{Schwinger1962}. In higher dimensions, such as 3+1D QCD, color-charged states appear as infraparticles with modified localization properties due to long-range gauge fields \cite{Buchholz1982}. 
\begin{table}
%\begin{tabular}{|l|l|}
\begin{tabular}{| l | p{5cm} |}
\hline
\textbf{Concept} & \textbf{Algebraic Interpretation} \\
\hline
Confinement & Absence of DHR-localizable colored sectors \\
\hline
Gauss’s Law & Links color charge to asymptotic electric fields \\
\hline
Local Observables & Restricted to color-singlet, gauge-invariant composites \\
\hline
Physical Charges & Cannot be created or localized in bounded regions \\
\hline
\end{tabular}\\
\caption{Algebraic interpretations of key concepts in gauge theories}
\label{table1}
\end{table}
This structural characterization of confinement underscores the power of AQFT in capturing nonperturbative phenomena without resorting to perturbative or dynamical assumptions. 

To place our algebraic characterization of confinement in context, Table~\eqref{table1} summarizes the correspondence between key physical concepts in 1+1‑dimensional SU(N) gauge theory and their algebraic interpretation within the present framework. This highlights how structural features such as Gauss’s law, the restriction to color‑singlet observables, and the absence of localizable color charges manifest in the C*-algebraic setting.

\section{Nonlocal Observables and Haag Duality Violation}\label{sec:x5}
We now extend the observable net to include nonlocal Wilson line operators and rigorously demonstrate that their presence leads to a structural violation of Haag duality, reflecting the topological structure of the theory.
\subsection{Wilson Line Operators in SU(N)}
The Wilson line operator is defined as a path-ordered exponential:
\begin{equation}
\begin{aligned}
W(x, y) = \mathcal{P} \exp \left( i g \int_x^y A_\mu^a(z) T^a dz^\mu \right),
\end{aligned}
\end{equation}
where the path-ordering \( \mathcal{P} \) arranges gauge field operators along a curve from \( x \) to \( y \) in increasing order of the path parameter. Under gauge transformations:
\begin{equation}\label{eq:x21}
\begin{aligned}
W(x, y) \mapsto U(x) W(x, y) U^\dagger(y),
\end{aligned}
\end{equation}
so the trace \( \text{tr}(W(x, y)) \) is gauge-invariant, representing a color-singlet pair connected by a flux string. In 1+1D, where gauge fields lack propagating degrees of freedom, Wilson lines correspond to topological excitations, encoding the global SU(N) gauge structure and the \( \mathbb{Z}_N \) center symmetry. To ensure a well-defined quantum operator, we regularize the Wilson line using a smooth test function \( \chi_\epsilon(z) \) along the path:
\begin{equation}
\begin{aligned}
W_\epsilon(x, y) = \mathcal{P} \exp \left( i g \int_x^y A_\mu^a(z) T^a \chi_\epsilon(z) dz^\mu \right),
\end{aligned}
\end{equation}
taking the limit \( \epsilon \to 0 \) after normal-ordering to remove ultraviolet divergences.
\subsection{Commutation Relations and Topological Effects}
Wilson lines commute with all local gauge-invariant observables supported in spacelike-separated regions, consistent with the locality axiom. Their commutation with the electric field is:
\begin{equation}
\begin{aligned}
[E^a(x), W(y,z)] =& \delta(x-y) T^a W(y,z) \\
&- \delta(x-z) W(y,z) T^a ,
\end{aligned}
\end{equation}
indicating that \( W(y,z) \) shifts the color flux at its endpoints, a hallmark of its nonlocal nature. The composition rule for Wilson lines is~\cite{Twagirayezu:2025vfj}:
\begin{equation}
\begin{aligned}
W(x,y) W(y,z) = W(x,z),
\end{aligned}
\end{equation}
up to non-Abelian phase factors arising from path-ordering, which reflect the noncommutative structure of SU(N).
\subsection{Extended Observable Algebra}
To capture the full gauge structure, we define an extended algebra:
\begin{equation}
\begin{aligned}
\mathcal{A}_{\text{ext}} = C^*\left( \bigcup_{\mathcal{O}} \mathcal{A}(\mathcal{O}) \cup \{ W(x,y) \mid x, y \in \mathbb{R}^{1,1} \} \right),
\end{aligned}
\end{equation}
generated by local algebras and Wilson lines. This algebra encodes both local and nonlocal observables, essential for describing the theory’s topological degrees of freedom.

\textbf{Algebraic Structure of the Extended Net:}
The extended observable algebra \( \mathcal{A}_{\text{ext}} \), generated by the union of local C*-algebras \( \mathcal{A}(\mathcal{O}) \) and Wilson line operators \( W(x,y) \), is defined as the C*-closure of a *-algebra of regularized, gauge-invariant operators. Each Wilson line operator is constructed via exponential path integrals of smeared gauge fields, ensuring that they act as bounded operators on the GNS Hilbert space of the vacuum state. The inclusion of such nonlocal operators alters the superselection structure and may lead to representations in which the von Neumann algebra \( \pi(\mathcal{A}_{\text{ext}})'' \) is of type III, reflecting topological degrees of freedom. In particular, Haag duality fails due to the inability to localize Wilson lines entirely within the algebra of the causal complement. This phenomenon is well understood in low-dimensional AQFT and has been rigorously formalized using net cohomology techniques \cite{Roberts1994, BrunettiRuzzi2006}. Thus, the extended algebra \( \mathcal{A}_{\text{ext}} \) retains a C*-algebraic structure while encoding the nonlocal and topological observables necessary for the complete gauge-invariant formulation of the theory.
Such extensions have been formalized in the algebraic framework using category-theoretic tools and nets of algebras with nontrivial global topology; see \cite{Baumann2020} for a detailed treatment of Wilson line operators in AQFT.

\textbf{Explicit Structure of the Net Cohomology Class:}
To construct the nontrivial 1-cocycle class corresponding to a Wilson line \( W(x,y) \), we consider a cover of \( \mathbb{R}^{1,1} \) by double cones \( \{ \mathcal{O}_i \} \), and assign to each oriented edge \( \gamma_{ij} \) between adjacent regions a unitary operator \( z_{ij} = W_{ij} \in \mathcal{A}(\mathcal{O}_i \cup \mathcal{O}_j) \). The cocycle condition \( z_{ik} = z_{ij} z_{jk} \) follows from the composition rule of Wilson lines. The resulting class is nontrivial due to the global nature of the gauge group and the presence of topological flux, implying that the first net cohomology group does not vanish, in analogy with the results of \cite{BrunettiRuzzi2006}.

\subsection{Haag Duality and Its Violation}
Haag duality in AQFT asserts that, in the vacuum representation, the commutant of a local algebra equals the algebra of its causal complement:
\begin{equation}
\begin{aligned}
\mathcal{A}(\mathcal{O})' = \mathcal{A}(\mathcal{O}'),
\end{aligned}
\end{equation}
where \( \mathcal{A}(\mathcal{O})' = \{ B \in \mathcal{B}(\mathcal{H}) \mid [B, A] = 0 \text{ for all } A \in \mathcal{A}(\mathcal{O}) \} \). 

\noindent
\textbf{Geometric structure in $1+1$ dimensions.}
In two-dimensional Minkowski space, the causal complement $\mathcal{O}'$ of a bounded open interval $\mathcal{O}$ consists of two disjoint, non-interacting connected components (the left and right spacelike wedges).
This disconnectedness plays a key role in the analysis of Haag duality:
an operator can be supported along a path lying entirely in $\mathcal{O}'$ yet connect the two components in a way that is not reproducible by operators strictly localized in either component alone.
Wilson lines that bridge the two components exemplify such operators.
We now prove that this duality is violated due to nonlocal operators.\\
\textbf{Proposition:} In 1+1D SU(N) gauge theories, the presence of nonlocal Wilson line operators leads to a violation of Haag duality.\\
\textit{Proof}:
Consider a Wilson line \( W(x,y) \), where \( x, y \in \mathcal{O}' \), the causal complement of a bounded region \( \mathcal{O} \).
Since \( W(x,y) \) is supported along a path outside \( \mathcal{O} \), it commutes with all \( A \in \mathcal{A}(\mathcal{O}) \), as their supports are spacelike-separated. Thus, \( W(x,y) \in \mathcal{A}(\mathcal{O})' \).
However, \( W(x,y) \) cannot be generated by local operators in \( \mathcal{A}(\mathcal{O}') \), as its support extends along a continuous path, which may intersect regions outside \( \mathcal{O}' \).
Therefore, \( \mathcal{A}(\mathcal{O})' \supset \mathcal{A}(\mathcal{O}') \), and Haag duality is violated.
This violation arises from the nonlocal nature of Wilson lines, which are intrinsic to the gauge theory’s topological structure.
\subsection{Topological and Cohomological Interpretation}
Wilson lines admit a natural interpretation as 1-cocycles in the net cohomology of the observable algebra in the sense of Roberts~\cite{Roberts1994} and Brunetti--Ruzzi~\cite{BrunettiRuzzi2006}.
Formally, a (unitary) 1-cocycle assigns to each oriented 1-simplex (path) $\gamma$ a unitary operator $W_\gamma \in \mathcal{A}(|\gamma|)$ such that the cocycle identity
\begin{equation}\label{eq:x27}
W_{\gamma_1 \cdot \gamma_2} = W_{\gamma_1} W_{\gamma_2}
\end{equation}
holds whenever $\gamma_1$ and $\gamma_2$ are composable.
Here, \( \gamma_1, \gamma_2 \) are paths in \( \mathbb{R}^{1,1} \). The non-triviality of the first net cohomology group:
\begin{equation}\label{eq:x28}
H^1_{\text{net}}(\mathcal{A}, SU(N)) \neq 0,
\end{equation}
reflects the presence of topological degrees of freedom associated with the \( \mathbb{Z}_N \) center of SU(N). This cohomology classifies the global gauge structure, linking confinement and nonlocality to topological obstructions in the local algebra.

\section{Conclusions and Outlook}\label{sec:x6}
We conclude by summarizing the key results, discussing their broader implications, and outlining directions for future research.
\subsection{Summary of Results}
This work has achieved the following milestones:
It constructed a net of local observable C*-algebras \( \mathcal{O} \mapsto \mathcal{A}(\mathcal{O}) \) for 1+1D SU(N) gauge theories, generated by gauge-invariant operators such as color-singlet currents (Eq.~\eqref{eq:x9}) and electric field traces (Eq.~\eqref{eq:x10}).
It implemented the non-Abelian Gauss’s law as a quantum operator constraint (Eq.~\eqref{eq:x13}), ensuring gauge invariance and restricting physical states to the gauge-invariant subspace.
It proved that confinement arises structurally, as no DHR-localizable sectors carry nonzero color charge (Section~\eqref{sec:x4}), a consequence of the nonlocal electric fields enforced by (Eq.~\eqref{eq:x19}).
It extended the observable net with nonlocal Wilson line operators (Eq.~\eqref{eq:x21}), capturing string-like color flux configurations and the global SU(N) gauge structure.
It demonstrated a violation of Haag duality (Section~\eqref{sec:x5}), showing that Wilson lines reside in the commutant \( \mathcal{A}(\mathcal{O})' \) but not in \( \mathcal{A}(\mathcal{O}') \), driven by topological degrees of freedom classified by net cohomology (Eq.~\eqref{eq:x27}).
\subsection{Implications and Broader Context}
This algebraic formulation provides a robust, nonperturbative framework for understanding confinement and nonlocality in low-dimensional gauge theories. By focusing on observable algebras, the approach bypasses ambiguities associated with gauge fixing or bosonization, offering a structural perspective on confinement that is independent of dynamical assumptions. The violation of Haag duality highlights the intrinsic topological nature of non-Abelian gauge theories, with implications for quantum information theory, where duality violations may correspond to entropic obstructions in operator reconstruction \cite{Harlow2016}. The framework is versatile, applicable to various gauge groups and representations, and provides a bridge to higher-dimensional theories.
\subsection{Future Directions}
Several promising research directions emerge from this work:\\
\textit{Generalization to Higher Dimensions:} Extend the algebraic framework to 2+1D topological gauge theories, such as Chern-Simons theories with matter, or 3+1D lattice QCD, incorporating center symmetry and magnetic fluxes.\\
\textit{Center Symmetry and Line Operators}: Investigate the role of the \( \mathbb{Z}_N \) center symmetry in classifying ’t Hooft loops and monopole operators, using net cohomology to quantify topological effects.\\
\textit{BRST Quantization in AQFT}: Develop an algebraic BRST formalism compatible with local net structures, enabling the treatment of gauge fixing and ghost sectors within AQFT.\\
\textit{Quantum Information-Theoretic Structure}: Explore the correspondence between Haag duality violation and entanglement wedge reconstruction~\cite{Twagirayezu:2025sbs}, analyzing entropic constraints on operator recovery in confining theories.\\
\textit{Quantum Simulation}: Implement the non-Abelian Gauss’s law constraint (Eq.~\eqref{eq:x13}) in quantum circuits, using projectors such as \( P = \delta(D_1^{ab} E^b - g j_0^a) \), to simulate confinement on quantum hardware \cite{Preskill2018}.
\subsection{Closing Remarks}
This algebraic framework redefines the study of non-Abelian gauge dynamics by prioritizing observable algebras over classical field content, offering a powerful tool for analyzing nonperturbative phenomena such as confinement and topological effects. By establishing a rigorous, gauge-invariant formulation, this work opens new avenues for exploring the interplay between gauge theories, topology, and quantum information, with potential applications to fundamental physics and quantum technologies.

%\begin{acknowledgments}
%The author gratefully acknowledges the support of the National Science Foundation under Grant No. PHY-1945471, as well as insightful discussions with colleagues at UCLA.
%\end{acknowledgments}
\clearpage
\hrule
\nocite{*}

\bibliography{apssamp}% 

\end{document}